\pdfoutput=1
\documentclass[english,manuscript]{aastex}
\usepackage[T1]{fontenc}
\usepackage[latin9]{inputenc}
\usepackage{amsbsy}
\usepackage{amstext}
\usepackage{graphicx}
\usepackage{esint}

\makeatletter

\shorttitle{Surface-shear shaped solar dynamo}
\shortauthors{Pipin \& Kosovichev}
\bibpunct[,]{(}{)}{;}{a}{,}{,}

\usepackage[english]{babel}

\usepackage{babel}

\makeatother

\usepackage{babel}
\begin{document}

\title{The mean-field solar dynamo with double cell meridional circulation
pattern}

\author{V.V. Pipin$^{1-3}$ and A.G. Kosovichev$^{3}$}

\affil{$^{1}$Institute of Solar-Terrestrial Physics, Russian Academy of
Sciences, \\
 $^{2}$ Institute of Geophysics and Planetary Physics, UCLA, Los
Angeles, CA 90065, USA \\
 $^{3}$Hansen Experimental Physics Laboratory, Stanford University,
Stanford, CA 94305, USA }
\begin{abstract}
Recent findings of helioseismology as well as advances in direct numerical
simulations of global dynamics of the Sun have indicated that in each
solar hemisphere the meridional circulation forms the two cells along
the in the convection zone. We investigate properties of a mean-field
solar dynamo with such double-cell meridional circulation. The dynamo
model also includes the realistic profile of solar differential rotation
(including the tachocline and subsurface shear layer), and takes into
account effects of turbulent pumping, anisotropic turbulent diffusivity,
and conservation of magnetic helicity. Contrary to previous flux-transport
dynamo models, it is found that the dynamo model can robustly reproduce
the basic properties of the solar magnetic cycles for a wide range
of model parameters and the circulation speed. The best agreement
with observations is achieved when the surface speed of meridional
circulation is about 12 m/s. For this circulation speed the simulated
sunspot activity shows good synchronization with the polar magnetic
fields. Such synchronization was indeed observed during the past sunspot
cycles 21 and 22. We compare theoretical and observed phase diagrams
of the sunspot number and the polar field strength and discuss the
peculiar properties of Cycle 23. 
\end{abstract}


\section{Introduction}

It is widely believed that the sunspot activity is governed by a hydromagnetic
turbulent dynamo operating deep in the solar convection zone. The
meridional circulation, which is observed on the surface as a steady
flow from the equator to the poles with a speed of 10-20 m/s, is often
considered as an important ingredient of the solar dynamo \citep{dc99,chrev05}.
It is suggested that the equator-ward meridional flow in the deep
convection zone can promote the drift of sunspot activity from mid
latitudes to the equator in the course of the 11-year solar cycle
(so-called ``butterfly diagram''), and that the near-surface pole-ward
flows can be responsible for the meridional drift of the poloidal
magnetic field from mid-latitudes to the poles. This idea is extensively
used in the Babcock-Leighton type solar dynamo models. Despite the
fact that, the mean-field dynamo models can reproduce the main properties
of the solar cycles without meridional circulation (e.g., \citealt{pk11,pi13r}),
the circulation affects details of the model. In particular, if the
dynamo effect is distributed in the bulk of the convection zone, the
circulation properties may affect the magnetic field distribution
with depth. For example, the meridional circulation pattern, which
is derived from a mean field model of the solar differential rotation
has the latitudinal flow concentrated to the boundaries of the convection
zone \citep{kit-r11}. Such circulation swaps the toroidal magnetic
fields to the bottom of the convection zone, and the poloidal magnetic
fields get concentrated to the poles \citep{pk11mf}. The helioseismology
inversions suggest that the global circulation may consist of two
cells in the radial direction (see, \citealt{mitrak07,zhao12}). In
the northern hemisphere, the flow direction is from the equator to
the poles at the top of the convection zone, and the flow direction
is reversed in the bottom cell. Such double-cell circulation is also
found in numerical simulations \citep{2006ApJ...641..618M,miesch11,guer2013}.
The influence of multi-cell circulation on the flux-transport dynamo
models was previously studied by \citet{2006MSAIS...9...71B} and
\citet{brun07}. They found that in the case of double-cell circulation
pattern, the evolution of the large-scale toroidal and poloidal magnetic
fields in such models can be qualitatively different from the solar
observations.

In this article we examine effects of the double-cell meridional circulation
pattern (see, Fig\ref{fig:profiles}d) in a mean-field dynamo model,
in which the magnetic field generation is distributed in the bulk
of the convection zone, and the migration of the magnetic field on
the surface is controlled by the subsurface rotational shear layer
action. This type of the subsurface-shear shaped dynamo was originally
suggested by \citet{2005ApJ...625..539B} and was extensively studied
in our recent papers (e.g., \citealt{pk11apjl}). In particular, it
was found that such dynamo model can reproduce the known statistical
relation of the solar cycles \citep{pipea2012AA}. Here, we investigate
effects of the double-cell meridional circulation pattern in this
model, and show that the resulted evolution of the large-scale magnetic
field is in good qualitative agreement with observations.

\section{The dynamo model}

We consider a large-scale axisymmetric magnetic field, $\bar{\mathbf{B}}=\mathbf{e}_{\phi}B+\nabla\times\frac{A\mathbf{e}_{\phi}}{r\sin\theta}$,
where $B(r,\theta,t)$ is the azimuthal component, $A(r,\theta,t)$
is proportional to the azimuthal component of the vector potential,
$r$ is the radial coordinate, and $\theta$ is the polar angle. The
mean flow is given by velocity vector $\mathbf{\overline{V}}=\mathbf{e}_{r}\overline{U}_{r}+\mathbf{e}_{\theta}\overline{U}_{\theta}+\mathbf{e}_{\phi}r\sin\theta\Omega$,
where $\Omega\left(r,\theta\right)$ is the angular velocity of the
solar differential rotation, and $\overline{U}_{r}(r,\theta)$ and
$\overline{U}_{\theta}(r,\theta)$ are velocity components of meridional
circulation. The mean magnetic field is governed by the induction
equation: 
\begin{equation}
\frac{\partial\overline{\mathbf{B}}}{\partial t}=\nabla\times\left(\mathbf{\overline{V}\times}\overline{\mathbf{B}}+\boldsymbol{\mathcal{E}}\right),\label{eq:mf}
\end{equation}
where the mean electromotive force, $\boldsymbol{\mathcal{E}}$, is
given by \citet{pi08Gafd}(hereafter P08). The model formulation is
essentially the same as in our previous papers \citep{pk11,pipea2012AA}.
The only difference is that we add the effect of the double-cell meridional
circulation in the dynamo equation. The mean electromotive force,
which describes effects of turbulent flows and turbulent magnetic
fields on the large-scale magnetic field evolution, is rather complicated.
The results of global dynamo simulations \citep{2011A&A...533A.108S,brown2011,bran2012AA,kap2012}
suggest that we have to take into account the full information about
$\boldsymbol{\mathcal{E}}$ to match the mean-field dynamo model to
the results of the direct numerical simulations. Below we are briefly
summarized the basic ingredients of the dynamo processes which are
included in the model. The further details can be found in Appendix.

\paragraph{\textit{Magnetic field generation.} }

The magnetic field generation effects in the model are due to the
differential rotation, turbulent kinetic helicity ($\alpha$-effect).
For the differential rotation, we take an analytical fit to the recent
helioseismology results of \citet{Howe2011JPh}, which include the
tachocline region at the bottom of the convection zone and the near-surface
rotational shear layer (see Figure 1c). The nonlinear $\alpha$-effect
which is computed on the base of the analytical expressions provided
by the mean-field theory, and with the help of the mixing-length estimations
for turbulent parameters from a standard solar interior model. In
addition we take into account the nonlinear effects of magnetic field
generation induced by the large-scale current and the global rotation,
which are usually called the $\Omega\times J$-effect or the $\delta$-effect
\citep{rad69,2008A&A...491..353K,2011A&A...533A.108S}. Figure 1a
shows the radial profiles of coefficients of the $\alpha$- and the
$\Omega\times J$- effects. One of the principal features of the model
is that it takes into account the subsurface shear layer. In addition
to the differential rotation in the bulk of the convection zone the
subsurface shear layer provides additional energy for the toroidal
magnetic field generation as well as it induces the equator-ward drift
of the toroidal magnetic field in the activity cycles.

\paragraph{\textit{Turbulent transport.} }

Turbulent transport of magnetic fields is related to effective drift
of large-scale magnetic fields in turbulent media in the absence of
mean flows. Theoretical models and global dynamo simulations suggest
that turbulent transport of large-scale magnetic fields in the solar
convection zone may result from several physical factors: the mean
density and turbulent intensity gradients (so-called ``gradient pumping'');
combined action of the large-scale vorticity and helicity (both kinetic
and magnetic helicity can contribute to the pumping); and effects
of global rotation on turbulent motions. The latter produces anisotropy
of the turbulent diffusivity and modifies the turbulent pumping effects.
It was found that these mechanisms are important for the latitudinal
migration of the dynamo wave in the solar convection zone \citep{k02,2003PhRvE..67b6321K,bran2012AA,garr2011}.

\paragraph{\textit{Meridional circulation.} }

The meridional flow is modeled in the form of four stationary circulation
cells with two cells along the radius in each hemisphere. The pattern
is modeled by the stream function $\Psi$, $\overline{\mathbf{U}}={\displaystyle \frac{U_{0}}{\overline{\rho}}\boldsymbol{\nabla}\times}\left(\mathbf{e_{\phi}\Psi}\right)$,
where we assume the following analytical form to describe the radial
and latitudinal dependence of $\Psi$: 
\begin{eqnarray}
\Psi & = & \frac{2c_{m}}{\pi}\left(x_{m}^{2}-x^{2}\right)\left(\frac{1}{x}-x_{e}\right)^{1.5}\sin\left(\frac{\pi\left(x-x_{i}\right)}{\left(x_{e}-x_{i}\right)}\right)\left(\frac{\partial P_{2}}{\partial\theta}+m\frac{\partial P_{4}}{\partial\theta}\right),\label{eq:psi}
\end{eqnarray}
where, $U_{0}$ is the amplitude of the flow, $x=r/R$ is a distance
in the units of the solar radius, $x_{m}=r_{m}/R$ is the stagnation
point of the upper cell, $x_{i}=r_{i}/R$ is the inner boundary of
the integration domain, $x_{e}=r_{e}/R$ is the upper boundary of
the integration domain; parameter $m$ controls the number of cells
in latitude; $c_{m}$ is the constant to normalize the maximum of
the flow amplitude to 1. In the model, we use $x_{m}=0.89$, and $m=-1/3$,
that gives $c_{m}=1/14$ for $x_{i}=0.715$ and $x_{e}=0.99$. The
given parameters result in the circulation pattern shown in Figure
1c. This pattern resembles qualitatively the results of the helioseismology
inversions (\citealp{mitrak07}, also Zhao et al 2012). Note, that
assuming $x_{m}=0$ and keeping the others parameters the same we
get a single-cell (per hemisphere) flow pattern which is close to
the mean-field models results (see, \citealt{kit-r11}). For $m>1/3$,
additional secondary latitudinal cells become pronounced at the high
latitude .

\paragraph{\textit{Conservation of magnetic helicity and dynamical $\alpha$-quenching}. }

The nonlinear feedback of the large-scale magnetic field to the $\alpha$-effect
is described by a dynamical quenching due to the constraint of the
total magnetic helicity conservation. The local helicity density is
the sum of the contributions from the small and the large-scale magnetic
fields: $\overline{\chi}^{(tot)}=\overline{\chi}+\overline{\mathbf{A}}\cdot\overline{\mathbf{B}}$
, where $\overline{\chi}=\overline{\mathbf{a}\cdot\mathbf{b}}$ ($\mathbf{a}$
and $\mathbf{b}$ are the fluctuating parts of magnetic field vector-potential
and magnetic field vector). The conservation of the total magnetic
helicity can be written as follows \citep{2012ApJ...748...51H,pi13r}:
\begin{eqnarray}
\frac{d}{dt}\int\overline{\chi}^{(tot)}dV & = & -\eta\int\left\{ \overline{\mathbf{B}}\cdot\mathbf{\overline{J}}+\overline{\mathbf{b\cdot j}}\right\} dV-\int\boldsymbol{\nabla\cdot}\boldsymbol{\boldsymbol{\mathcal{F}}}^{\chi}dV\label{eq:int-cons}
\end{eqnarray}
where integration is performed over the volume that comprises the
dynamo region; $\boldsymbol{\boldsymbol{\mathcal{F}}}^{\chi}$ is
the diffusive flux of the total helicity, which results from the turbulent
motions\citep{mitra10}, $\eta$ is the coefficient of molecular diffusivity.
The differential equation that corresponds to Eq.(\ref{eq:int-cons})
is:

\begin{equation}
\frac{\partial\overline{\chi}^{(tot)}}{\partial t}=-\frac{\overline{\chi}}{R_{m}\tau_{c}}-\eta\overline{\mathbf{B}}\cdot\mathbf{\overline{J}}-\boldsymbol{\nabla\cdot}\boldsymbol{\boldsymbol{\mathcal{F}}}^{\chi}-\mathbf{\left(\overline{U}\cdot\boldsymbol{\nabla}\right)}\overline{\chi}^{(tot)}\label{eq:helcon}
\end{equation}
where ${\boldsymbol{\mathcal{F}}}^{\chi}=-\eta_{\chi}\boldsymbol{\nabla}\overline{\chi}^{(tot)}$.
In the model we assume $R_{m}=10^{6}$ and $\eta_{\chi}=0.1\eta_{T}$.
The magnetic helicity contribution to the $\alpha$-effect is defined
as follows (P08): 
\begin{equation}
\alpha_{ij}^{(M)}=2\left(f_{2}^{(a)}\delta_{ij}-f_{1}^{(a)}\frac{\Omega_{i}\Omega_{j}}{\Omega^{2}}\right)\frac{\overline{\chi}\tau_{c}}{\mu_{0}\overline{\rho}\ell^{2}}\label{alpM}
\end{equation}

The mixing-length is defined as $\ell=\alpha_{{\rm MLT}}\left|\Lambda^{(p)}\right|^{-1}$,
where $\Lambda{}^{(p)}=\boldsymbol{\nabla}\log\overline{p}\,$ is
the inverse pressure scale height, and $\alpha_{{\rm MLT}}=2$. We
use the solar convection zone model of \citet{stix:02}. Since the
mixing-length theory provides approximate estimates of the turbulent
properties we introduced scaling factors, and investigate effects
of scaling on the dynamo regimes. The turbulent diffusivity is parameterized
in the form, $\eta_{T}=C_{\eta}\eta_{T}^{(0)}$, where $\eta_{T}^{(0)}={\displaystyle \frac{u'^{2}\tau_{c}}{3f_{ov}\left(r\right)}}$
is a profile of the mixing-length turbulent diffusivity, $\ell$ is
a typical correlation length of the turbulence; $C_{\eta}$ is a constant
to control the efficiency of large-scale magnetic field dragging by
the turbulent flow; it determines the period of the magnetic cycles.
Also, we modify the mixing-length turbulent diffusivity by factor
$f_{ov}(r)=1+\exp\left(50\left(r_{ov}-r\right)\right)$, where $r_{ov}=0.725R_{\odot}$,
which models saturation of the turbulent parameters near the bottom
of the convection zone, as suggested by the numerical simulations
(see, e.g., \citealp{1995A&A...296..557R,oss2001,oss02,2008A&A...491..353K}).
The results do not change much if we scale the turbulent diffusivity
gradient $\mathbf{\Lambda}^{(u)}=C_{v}\boldsymbol{\nabla}\log\left(\eta_{T}^{(0)}\right)$
with a factor $C_{v}\le0.5$. For greater values of $C_{v}$ our model
leads to a steady non-oscillating dynamo which is concentrated to
the bottom of the convection zone. Note, that the previous flux-transport
models \citep{2002A&A...390..673B,2006MSAIS...9...71B} use $C_{v}=0$. 

\begin{figure}
\begin{centering}
\includegraphics[width=0.99\columnwidth]{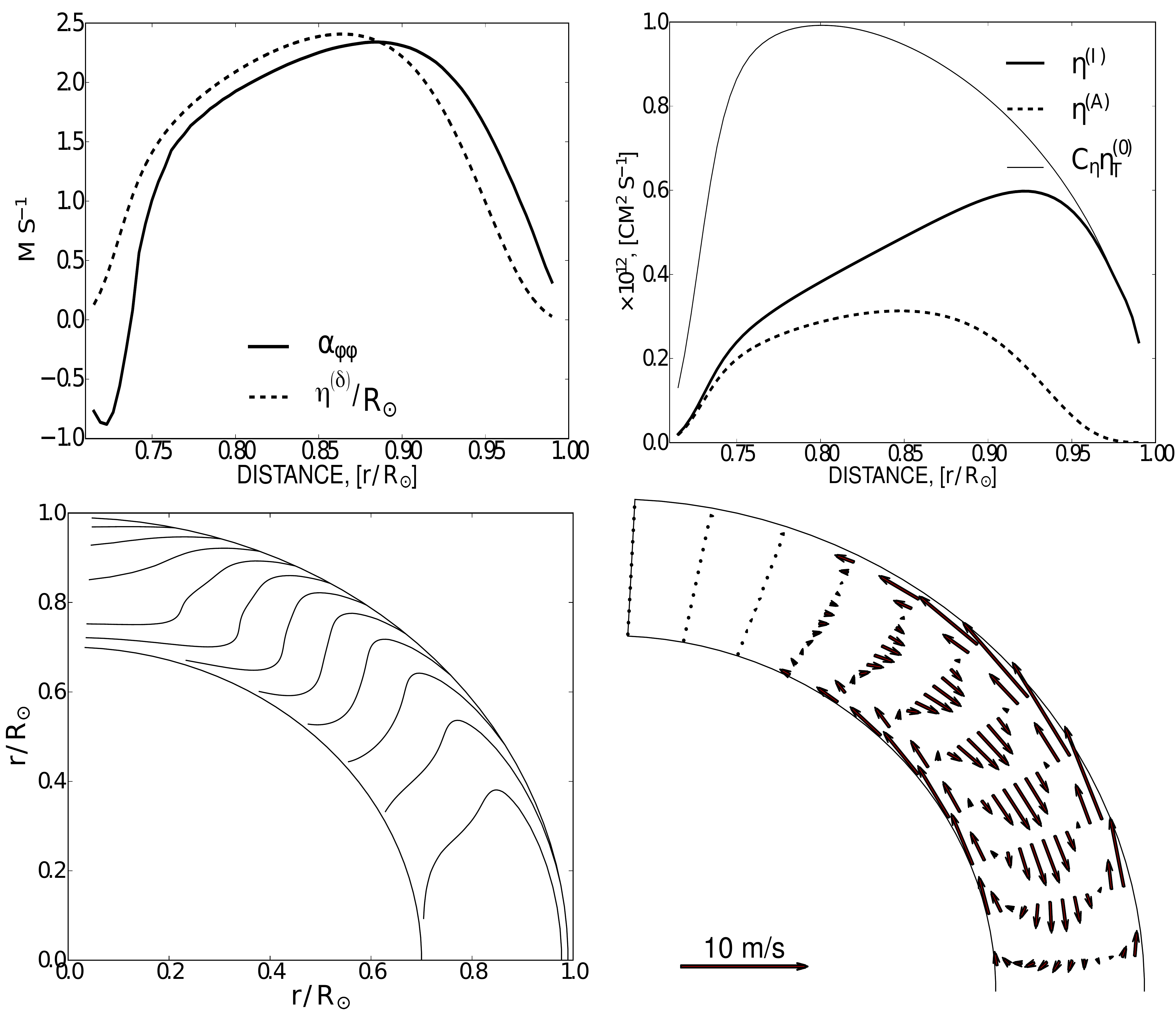} 
\par\end{centering}

\caption{Panel (a), the radial profiles of components of the helicity coefficient,
$\alpha$, (Eq.\ref{alp2d}) and the $\Omega\times J$ term (Eq. \ref{eq:delta})
for $\theta=45^{\circ}$, the $\Omega\times J$ effect was normalized
by $R_{\odot}$ and amplified by factor 10 to show at the same amplitude
range as the $\alpha$ effect; panel (b), the radial profiles of the
background turbulent diffusivity $C_{\eta}\eta_{T}^{(0)}$, the isotropic,
$\eta^{(I)}$, and anisotropic, $\eta^{(A)}$, parts of the magnetic
diffusivity coefficients;c) the angular velocity distribution; (d)
velocity vectors of the double-cell meridional circulation pattern
\label{fig:profiles}.}
\end{figure}

The bottom of the integration domain is at $r_{b}=0.715R_{\odot}$,
and the top of the integration domain is $r_{e}=0.99R_{\odot}$. \citet{pk11mf}
showed that the solar-type dynamo solution can be obtained for $C_{\alpha}/C_{\delta}\ge2$.
In that paper we found that the approximate threshold for magnetic
field generation is $C_{\alpha}\approx0.03$ for a diffusivity scaling
factor $C_{\eta}=0.08$ to match the 22-year periodicity. Figure \ref{fig:profiles}
shows the radial profiles for the $\alpha$-effect coefficients, the
isotropic and anisotropic parts of the turbulent diffusivity and the
$\Omega\times J$ effect. They are in the qualitative agreement with
the simulation results by \citet{oss2001} and \citet{2009A&A...500..633K}.

For axisymmetric large-scale magnetic fields the vector-potential
can be represented as a sum of the azimuthal and poloidal components
\citep{krarad80}: 
\begin{equation}
\overline{\mathbf{A}}=\mathbf{e}_{\phi}T+\mathbf{r}P=\frac{\mathbf{e}_{\phi}}{r\sin\theta}A+r\mathbf{e}_{r}P.\label{eq:potent}
\end{equation}
The toroidal part of the vector potential, $A$, is governed by the
dynamo equations (Eq.\ref{eq:mf}). The poloidal part of the vector
potential, $P$, can be found from the toroidal magnetic field component,
$B$, using equation $\boldsymbol{\nabla}\times\left(\mathbf{r}P\right)=\mathbf{e}_{\phi}B$.
The reconstruction is simple by using pseudo-spectral numerical schemes
which are based on the Legendre polynomial decomposition for the latitudinal
profile of $B$. Following \citet{pk11apjl} we use a combination
of the ``open'' and ``closed'' boundary conditions at the top,
controlled by a parameter $\delta=0.99$: 
\begin{equation}
\delta\frac{\eta_{T}}{r_{e}}B+\left(1-\delta\right)\mathcal{E}_{\theta}=0.\label{eq:tor-vac}
\end{equation}
This is similar to the boundary condition discussed by \citet{kit:00}.
For the poloidal field we apply a combination of the local condition
$A=0$ and condition of smooth transition from the internal poloidal
field to the external potential (vacuum) field: 
\begin{equation}
\delta\left(\left.\frac{\partial A}{\partial r}\right|_{r=r_{e}}-\left.\frac{\partial A^{(vac)}}{\partial r}\right|_{r=r_{e}}\right)+\left(1-\delta\right)A=0,\label{eq:pol-vac}
\end{equation}
For the magnetic helicity we employ $\bar{\chi}=0$ at the bottom
of the convection zone, and we assume that the radial derivative of
the total helicity is zero at the top. The initial magnetic field
is assumed dipolar with a small quadrupolar component to check the
parity preference when the solution reaches a stationary state.

\section{Results and discussion}

\begin{figure}
\begin{centering}
\includegraphics[width=0.4\textwidth]{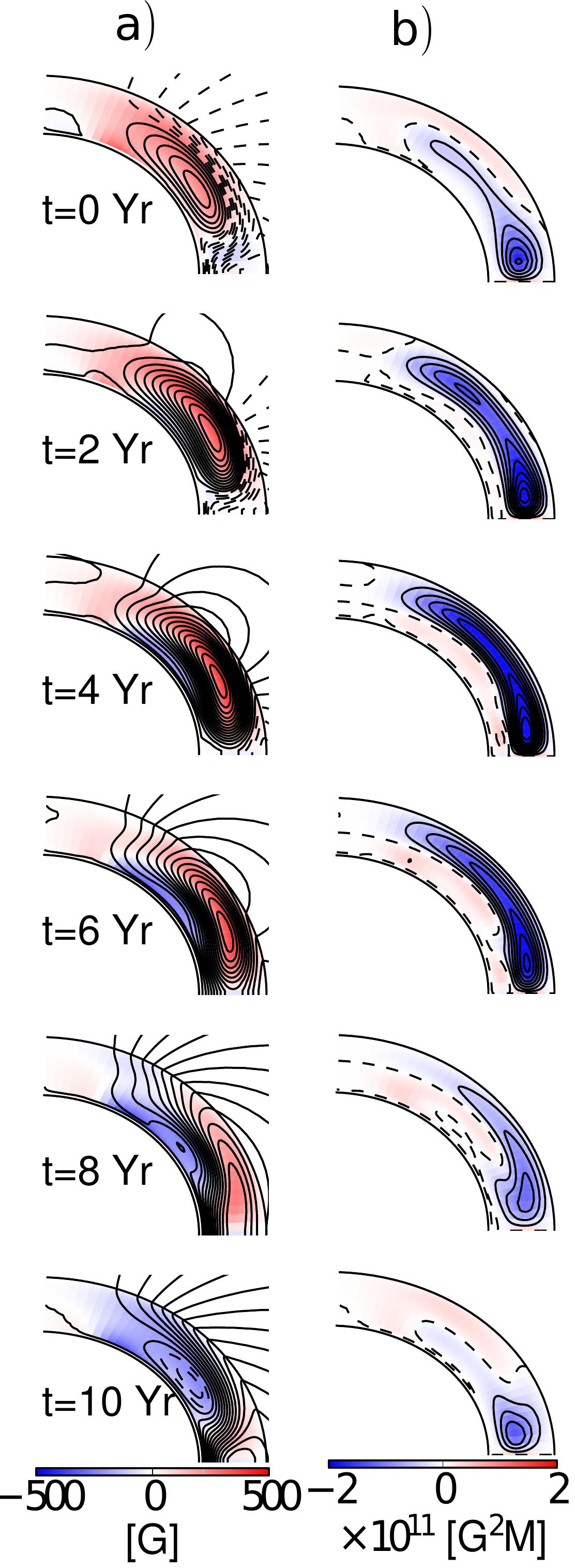}
\par\end{centering}

\caption{a) Evolution of the large-scale magnetic field inside the convection
zone for the dynamo model with the meridional circulation speed $U_{0}=12$
m/s. The field lines show of the poloidal component of the mean magnetic
field, and the toroidal magnetic field (varies $\pm0.6$kG) is shown
by the background images. b) Evolution of the large-scale magnetic
helicity density (contours) and the mean helicity density of small-scale
fields (background images). Both parts of the helicity density vary
within the same range of magnitudes.\label{fig:Snapshots}}
\end{figure}

Figure \ref{fig:Snapshots} shows the snapshots of the magnetic field
and magnetic helicity evolution in the North segment of the solar
convection zone for the model with the circulation speed $U_{0}=12$m/s.
The similar snapshots for the model without circulation can be found
in \citep{pip2013ApJ}. We observe drift of the dynamo waves related
to the evolution of the large-scale toroidal and poloidal fields,
towards the equator and towards the pole, respectively. It is found
that the meridional circulation accelerates the equator-ward drift
of the toroidal magnetic field near the surface and the poleward drift
of the toroidal magnetic field near the bottom of the convection zone. 

The snapshots on Figure \ref{fig:Snapshots} illustrate the evolution
of the large-scale magnetic fields (Fig.2a) and magnetic helicity
(Fig.2b). The toroidal magnetic field of the new cycle is generated
near the bottom of the convection zone by the differential rotation.
Simultaneously, we see a start of generation of the poloidal magnetic
field (contour lines) by the $\alpha$- and $\Omega\times J$ effects.
The dynamo wave propagates by the turbulent diffusion processes almost
radially to the surface following the Parker-Yoshimura rule \citep{1955ApJ...121..491P,yosh1975}.
However, the propagation of the wave is inclined to the equator because
of the anisotropy of the turbulent diffusion and turbulent transport
effects \citep{k02}. Near the surface the turbulent downward pumping
and the subsurface rotational shear \citep{pk11} stop the radial
propagation and deflect the dynamo wave toward the equator. The near-surface
meridional circulation and the turbulent diffusion bring the decaying
poloidal field to the poles. 

The meridional circulation modifies the propagation of the dynamo
wave. It is found that the toroidal magnetic field is strongly involved
in clockwise advection by the bottom circulation cell in a manner
which is typical for the flux transport models. The inner part of
the poloidal magnetic field flux is carried by the meridional flow
in a similar way. Near the surface the poloidal field migrates towards
the poles at high latitudes and towards the equator at low latitudes. 

Figure \ref{fig:Snapshots}b shows the evolution of large-scale (contours)
and small-scale (colors) magnetic helicities. As suggested by the
helicity conservation law (Eq. \ref{eq:helcon}), the evolution of
the mean helicity density of small-scale fields follows the evolution
of the helicity density of large-scale magnetic fields. In the upper
part of the convection zone the small-scale magnetic helicity obeys
the so-called helicity hemispheric sign rule (hereafter ``magnetic
helicity sign rule'') \citep{see1994AA}: it is negative sign in
the Northern hemisphere and positive in the Southern hemisphere. The
reversals of the helicity hemispheric rule can be found at high latitudes
at the beginning of the cycle when the large-scale magnetic helicity
is determined by the contribution of the near surface poloidal magnetic
field (cf snapshots for the $t=2$yr). The reversals of the magnetic
helicity sign rule have the different origin in the flux-transport
and mean-field dynamo models (see, e.g, \citealp{choud2004ApJ,2012ApJ...751...47Z,pi13r,pip2013ApJ}).

\begin{figure}
\includegraphics[width=0.99\textwidth]{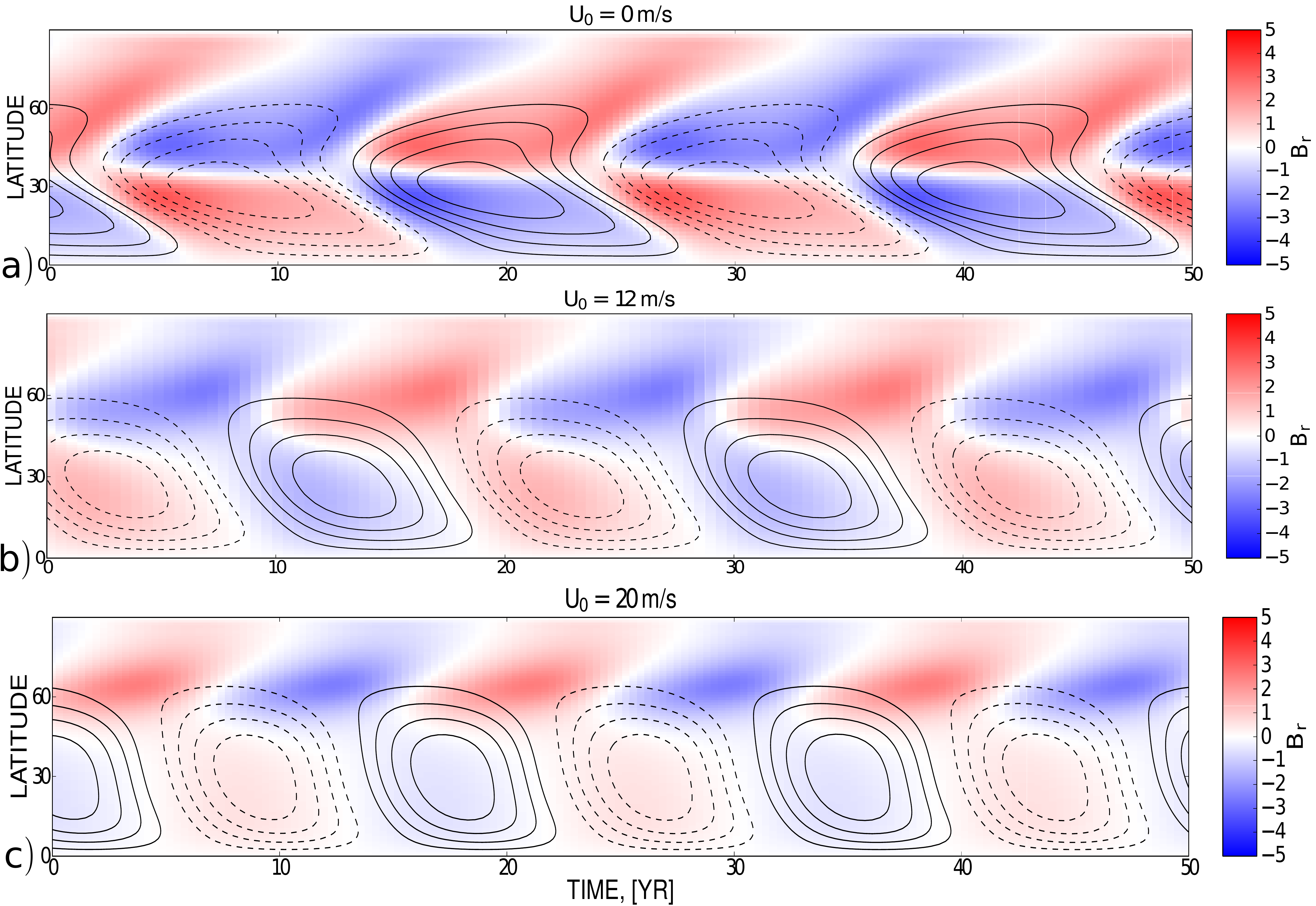}

\caption{(a) the time - latitude variations (``butterfly diagrams'') for
the model without meridional circulation; (b) the same for the model
with the meridional circulation of $U_{0}=12$ m/s; (c) the same as
(b) for $U_{0}=20$ m/s. The toroidal field near the surface, at $r=0.92R_{\odot}$,
is shown by contours (plotted for $\pm100$G range), and the surface
radial magnetic field is shown by background red-blue images.\label{fig:diagr}}
\end{figure}

Figure \ref{fig:diagr} shows the time-latitude ``butterfly'' diagrams
of the toroidal (contours) and radial (colors) magnetic fields evolution
in the upper part of the solar convection zone for the meridional
flow speed $U_{0}=0$, 12 and 20 m/s. In all cases (with and without
the circulation) there is the qualitative agreement with observations.
We found that the maximum of the toroidal magnetic field migrates
closer to the equator for the models which include the circulation.
However, in these cases the butterfly diagram wings are wider in latitude
than in the case without circulation. Also it is found that the circulation
reduces the latitudinal width of the polar branch for the radial magnetic
field evolution and also reduces the overlap between the cycles. We
draw these diagrams only for one hemisphere because in all these cases
the antisymmetric mode (dipole-like) is dominant.

\begin{figure}
\includegraphics[width=0.99\textwidth]{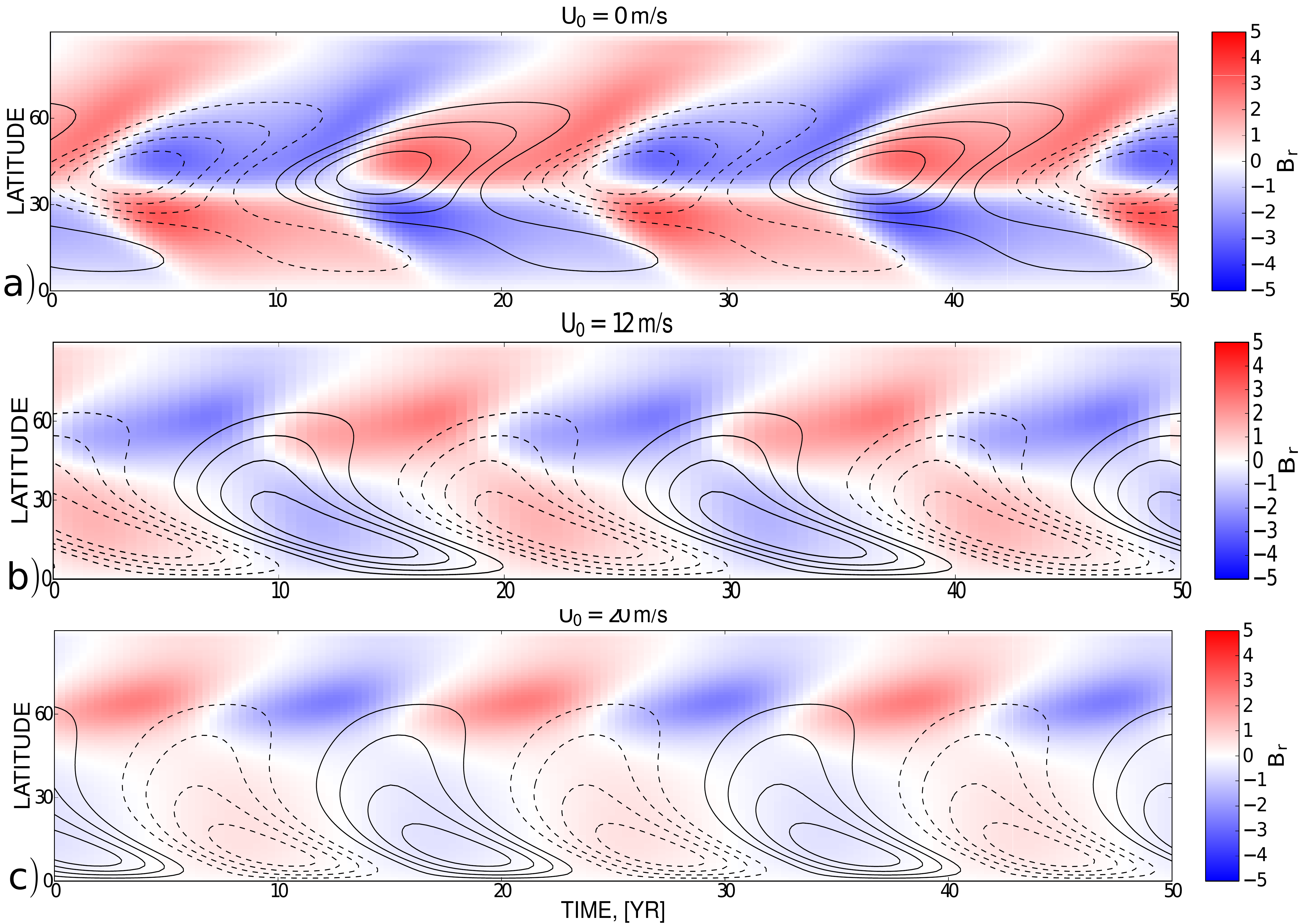}

\caption{The same as Figure \ref{fig:diagr} for the toroidal magnetic field
(contour lines) at $r=0.82R_{\odot}$.\label{fig:diagr-1}}
\end{figure}

Figure \ref{fig:diagr-1} shows the toroidal magnetic field evolution
in the middle part of the solar convection zone where the flows of
the two cells converge. In the model without circulation the polar
branch of the toroidal magnetic field is dominant. When the amplitude
of the circulation speed is greater than $U_{0}=8$m/s the toroidal
magnetic field is swept towards the equator. In this case the time-latitude
patterns of the toroidal magnetic field evolution in the middle of
the convection zone are very similar to those suggested by the standard
flux-transport and advection dominated dynamos \citep{2002A&A...390..673B,2006ApJ...647..662R,brun07}
near the bottom of the convection zone.

\begin{figure}
\includegraphics[width=0.95\linewidth]{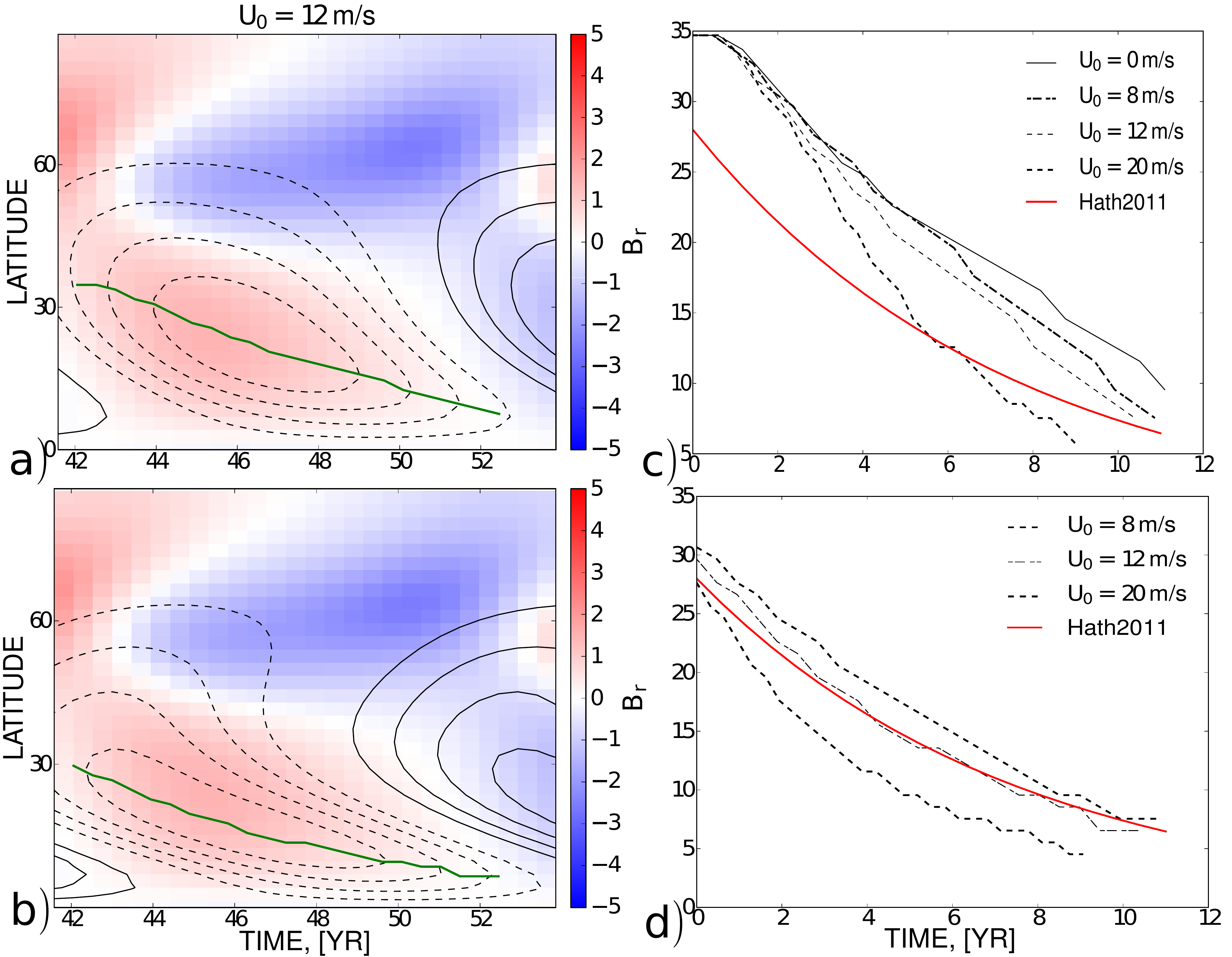}

\caption{To calculation of the latitudinal coordinates of the maximum of the
toroidal magnetic field flux. Panels a) and b) show the examples of
the time-latitude pattern of the magnetic field evolution at $r=0.92R_{\odot}$(a)
and $0.82R_{\odot}$(b) for the model with $U_{0}=12$m/s. The toroidal
field is shown by contours, and the radial field is shown by background
images. The green solid line shows position of the toroidal magnetic
field flux maxima. Panels c) and d) show the results of calculations
for the different models at $r=0.92R_{\odot}$(c) and $0.82R_{\odot}$(d)
\label{fig:centroid}. Red line marks the observational results by
\citet{2011SoPh..273..221H}.}
\end{figure}

To compare the butterfly diagram shapes with solar observations we
calculate the latitudinal coordinates of the maximum of the toroidal
magnetic field flux. The results for the magnetic flux that resides
at two depths ($r=0.92R_{\odot}$ and $r=0.82R_{\odot}$) in the solar
convection zone are shown in the Figure \ref{fig:centroid}. The observations,
(see e.g., \citealp{2011SoPh..273..221H}) suggest that the latitude
of the maximum of the sunspot formation decreases exponentially in
course of the solar cycle. The model with $U_{0}=12$m/s is in the
best agreement with the observations if the sunspot flux originates
at $r=0.82R_{\odot}$. In the model, which does not include circulation
the toroidal magnetic field flux does not approach the equator as
close as in the model with circulations, which show somewhat better
agreement with the observations. However their qualitative behaviors
are similar. 

\begin{figure}

\includegraphics[width=0.6\columnwidth]{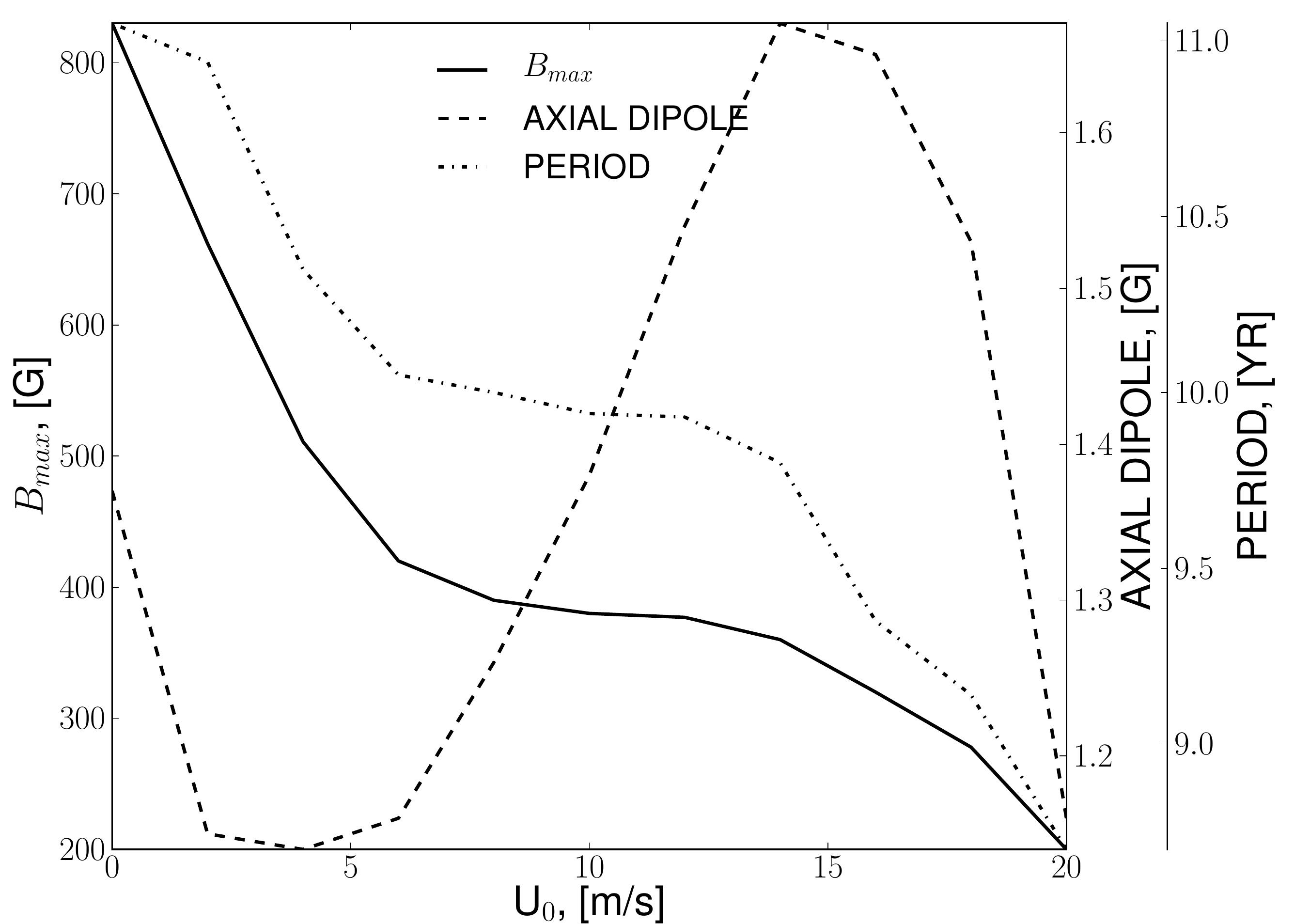}\caption{Dependences of the maximum of the toroidal magnetic field strength
in the convection zone, the dynamo period and the magnitude of the
axial dipole on the amplitude of the meridional circulation speed,
$U_{0}$.\label{fig:u-bmax}}

\end{figure}

Figure \ref{fig:u-bmax} shows the dependence of maximum of the toroidal
magnetic field strength inside the solar convection zone, the dynamo
period and the magnitude of the axial magnetic dipole on the amplitude
of the meridional circulation $U_{0}$. All parameters were computed
for the same dynamo parameter $C_{\alpha}$. The amplitude of the
generated magnetic field as well as the dynamo period decrease monotonically
with the increase of the circulation speed. However, there is a plateau
in the range of $U_{0}$ between 8m/s and 14m/s. Interesting that
the magnitude of the generated axial dipole varies substantially within
the same range. The growth of the magnetic dipole corresponds to the
enhanced poloidal magnetic field generation in this interval of the
circulation speed, which can be due to a resonance between the meridional
circulation and the dynamo wave propagation in the middle of the convection
zone. This suggestion requires further detailed studies. 

\begin{figure}
\includegraphics[width=0.95\columnwidth]{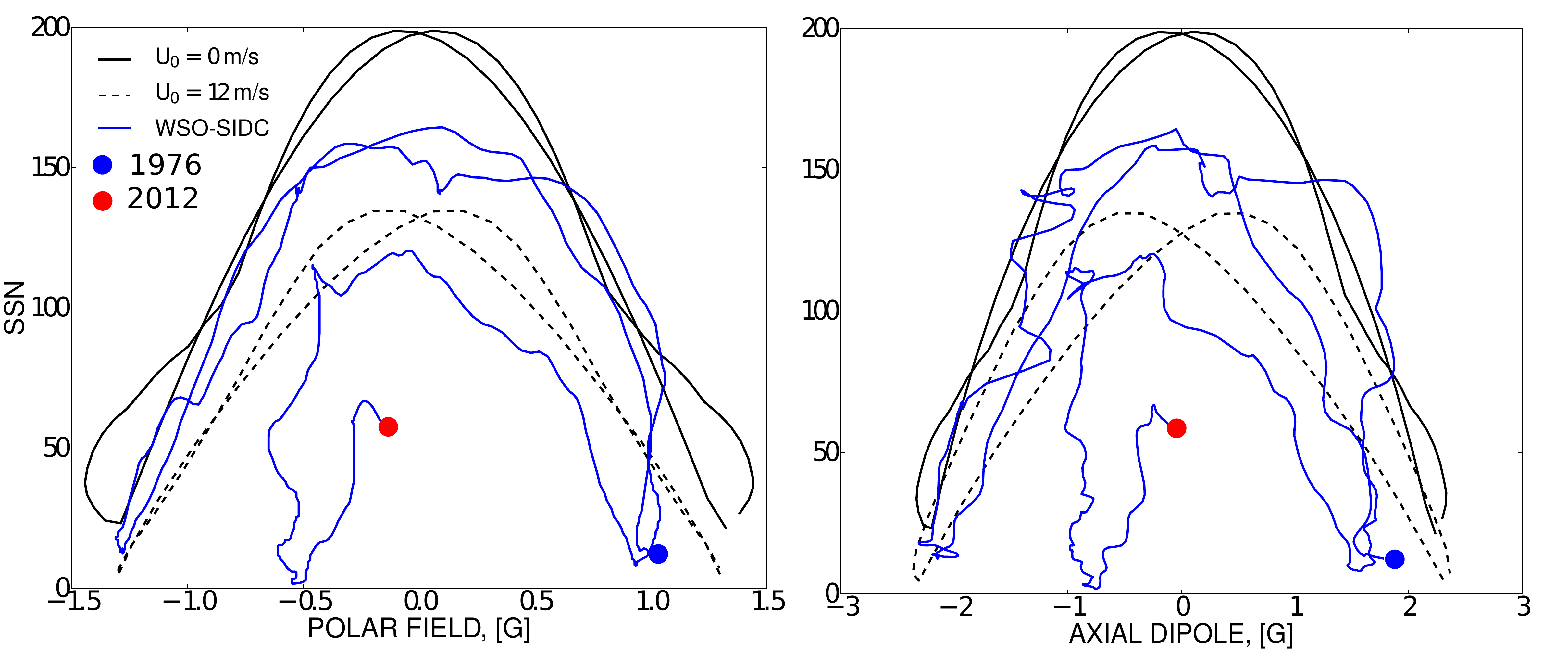}\caption{The phase relations between a) the strength of the polar magnetic
field and the sunspot number and b) the axial dipole and the sunspot
number in our models (black curves) and in the observations (noisy
blue curves).\label{fig:phase} }
\end{figure}

The interesting question is how the circulation changes the phase
relation between the polar magnetic field strength and the sunspot
activity in our model. The simulated sunspot number (SSN) was computed
following \citet{pipea2012AA}, as $\textnormal{SSN}={\displaystyle B_{max}\exp\left(-\frac{B_{0}}{B_{max}}\right)}$,
where $B_{0}=600$G and $B_{max}$ is the maximum of the toroidal
magnetic field strength at $r=0.92R_{\odot}$. The results are shown
in Figure \ref{fig:phase}. To compare with observations we used the
smoothed sunspot number provided by SIDC \citep{sidc}. The observations
of the Wilcox Solar Observatory \citep{setal78,hoek95} provide us
with the strength of the line-of-sight polar magnetic field which
was averaged over the polemost $3^{\prime}$ aperture. The magnitude
of the axial dipole was computed from the coefficients of the potential
extrapolation of the surface magnetic field. In the model we will
assume that the measured signal of the line-of-sight component of
the field is formed by the radial component of the large-scale magnetic
field. The strength of the axial dipole in the model is calculated
from the first coefficient of the vector-potential expansion at the
top boundary. Figure \ref{fig:phase} illustrates the phase relations
between the sunspot number and the polar radial magnetic field in
the models and in observations. In the phase digrams the maxima of
the polar magnetic field and the axial dipole correspond to the minima
of the sunspot number. The reversals of the polar magnetic field correspond
to the sunspot maxima. For the axial dipole this phase relation fluctuates
and observations show a delay of about two years in the last solar
cycle, which is not explained in our models.

In the case of ideal synchronization the two subsequent cycles would
produce the parabolic curve in the phase diagrams. This is also found
for the sunspot Cycles 21 and 22 which were not very different in
amplitude. Cycle 23 had a smaller magnitude, and started with the
polar magnetic field of about 1 G in the minimum as the previous two
cycles. While our model qualitatively reproduces the phase relation,
it does not explain observed a sudden decrease of the sunspot activity
in Cycle 23. 

In these calculations, we assumed that the sunspot activity originates
from the toroidal magnetic field at $r=0.92R_{\odot}$ ( just below
the subsurface shear layer) because this depth provided the best agreement
with observations for the model without circulation flow. It is interesting
that for the models with the meridional flow the acceptable range
of the depths is significantly broader.

We investigated the stability of the obtained solutions with respect
to changes of the circulation pattern and the other parameters of
the model. The results similar shown in Figures 2 and 3, are obtained
for the case of pure a $\alpha^{2}\Omega$ dynamo (without the $\Omega\times J$
effect). However, in this case, the cycle overlap in the time-latitude
diagrams is much stronger than in Figures 3(b,c). The similar time-latitude
butterfly diagrams are obtained also for the case when the meridional
circulation forms only one cell per hemisphere with the stagnation
point radius larger than $0.85R$. This is because of the subsurface
rotational shear layer, which plays the primary role in the model.
For the deeper stagnation point the dynamo wave is dominated by the
toroidal magnetic field near the bottom of the convection zone \citep{pk11mf}.
Also, we checked the effect of the meridional circulation in the simplified
model which was presented by \citet{pk11apjl}. In that model the
$\alpha$-effect was confined in the low-latitude $\pm30^{\circ}$
zone. The results are generally similar to what is shown in Figures
2 and 3.

\section{Conclusions}

The main findings can be summarized as follows. The mean-field dynamo
model that includes the subsurface rotational shear layer and the
double-cell (in radius) meridional circulation, indicated by the recent
helioseismology results, can reproduce the solar magnetic cycles in
the form of the time-latitude ``butterfly'' diagrams. The double-cell
circulation affects the distribution of the magnetic field with radius
in the convection zone, increasing the field concentration to the
convection zone boundaries, and n the middle of the convection zone
where the two cells converge. The latter effect can lead to a non-monotonic
profile of the amplitude of the large-scale poloidal magnetic field
in response to an increase of the circulation speed. The models qualitatively
explains the observed synchronization between the polar magnetic field
strength and the sunspot number. However, it does not explain the
sudden drop of the Sun's magnetic activity in Cycle 23. This question
has to be examined further in order to clarify the detail of this
synchronization. By comparing the toroidal field migration speed at
the different depths with the observed characteristics speed of the
sunspot formation zone we found that the range of the depth of the
active region emerging flux can be between 0.82-0.92$R_{\odot}$.
This estimate can be improved for a more precise helioseismology measurements
of the meridional circulation.

\section{Appendix}

Here, we give some details about calculations of the mean electromotive
force. The more elaborated description can be found in one of our
previous papers (e.g., \citealp{pk11}). The mean electromotive force
is given as follows  (Pipin, 2008). 
\begin{equation}
\mathcal{E}_{i}=\left(\alpha_{ij}+\gamma_{ij}^{(\Lambda)}\right)\overline{B}_{j}-\left(\eta_{ijk}+\eta_{ijk}^{(\delta)}\right)\nabla_{j}\overline{B}_{k}.\label{eq:EMF-1}
\end{equation}
The tensor $\alpha_{ij}$, representing the so-called $\alpha$-effect,
includes hydrodynamic ($\alpha_{ij}^{(H)}$) and magnetic ($\alpha_{ij}^{(M)}$)
helicity contributions: 
\begin{eqnarray}
\alpha_{ij} & = & C_{\alpha}\psi_{\alpha}\sin^{2}\theta\alpha_{ij}^{(H)}+\alpha_{ij}^{(M)},\label{alp2d}\\
\alpha_{ij}^{(H)} & = & 3\eta_{T}\frac{\delta_{ij}}{\Omega}\left\{ \left(f_{10}^{(a)}\left(\boldsymbol{\mathbf{\Omega}}\cdot\boldsymbol{\Lambda}^{(\rho)}\right)+f_{11}^{(a)}\left(\boldsymbol{\mathbf{\Omega}}\cdot\boldsymbol{\Lambda}^{(u)}\right)\right)\right\} \\
 & + & 3\eta_{T}\frac{\Omega_{i}\Omega_{j}}{\Omega^{3}}\left\{ \left(f_{5}^{(a)}\left(\boldsymbol{\mathbf{\Omega}}\cdot\boldsymbol{\Lambda}^{(\rho)}\right)+f_{4}^{(a)}\left(\boldsymbol{\mathbf{\Omega}}\cdot\boldsymbol{\Lambda}^{(u)}\right)\right)\right\} \nonumber \\
 & + & \frac{3\eta_{T}}{\Omega}\left\{ \left(\Omega_{i}\Lambda_{j}^{(\rho)}+\Omega_{j}\Lambda_{i}^{(\rho)}\right)f_{6}^{(a)}+\left(\Omega_{i}\Lambda_{j}^{(u)}+\Omega_{j}\Lambda_{i}^{(u)}\right)f_{8}^{(a)}\right\} ,
\end{eqnarray}
where $\mathbf{\Lambda}^{(\rho)}=\boldsymbol{\nabla}\log\overline{\rho}$
quantifies the density stratification (inverse density scale height),
$\mathbf{\Lambda}^{(u)}=\frac{1}{2}\boldsymbol{\nabla}\log\left(\eta_{T}^{(0)}\right)$
quantifies the stratification of turbulent diffusivity $\eta_{T}^{(0)}$.
The $\alpha$-quenching function $\psi_{\alpha}=-3/4\phi_{6}^{(a)}\left(\beta\right)$
depends on $\beta={\displaystyle \left|\overline{B}\right|/\sqrt{\overline{\rho}\mu_{0}\overline{u^{2}}}}$,
and $\phi_{6}^{(a)}$ is given in P08.

The turbulent pumping coefficient, $\gamma_{ij}^{(\Lambda)}$, depends
on the mean density and turbulent diffusivity stratification, and
also on the Coriolis number $\Omega^{*}=2\tau_{c}\Omega_{0}$, where
$\tau_{c}$ is a typical convective turnover time, and $\Omega_{0}$
is the angular velocity. Following the results of P08, $\gamma_{ij}^{(\Lambda)}$
is expressed as follows: 
\begin{eqnarray}
\gamma_{ij}^{(\Lambda)} & = & 3\eta_{T}\left\{ f_{3}^{(a)}\Lambda_{n}^{(\rho)}+f_{1}^{(a)}\left(\boldsymbol{\Omega}\cdot\boldsymbol{\Lambda}^{(\rho)}\right)\frac{\Omega_{n}}{\Omega}\right\} \varepsilon_{inj}-3\eta_{T}f_{1}^{(a)}\frac{\Omega_{j}}{\Omega^{2}}\varepsilon_{inm}\Omega_{n}\Lambda_{m}^{(\rho)},\label{eq:pump}\\
 & - & 3\eta_{T}\left(\varepsilon-1\right)\left\{ f_{2}^{(a)}\Lambda_{n}^{(u)}+f_{1}^{(a)}\frac{\left(\boldsymbol{\Omega}\cdot\boldsymbol{\Lambda}^{(u)}\right)}{\Omega^{2}}\Omega_{n}\right\} \varepsilon_{inj}.\nonumber 
\end{eqnarray}
The turbulent diffusivity is anisotropic due to the Coriolis force,
and is given by: 
\begin{equation}
\eta_{ijk}=3\eta_{T}\left\{ \left(2f_{1}^{(a)}-f_{2}^{(d)}\right)\varepsilon_{ijk}-2f_{1}^{(a)}\frac{\Omega_{i}\Omega_{n}}{\Omega^{2}}\varepsilon_{njk}\right\} .\label{eq:diff}
\end{equation}
We also include the nonlinear effects of magnetic field generation
induced by the large-scale current and global rotation, which are
usually called the $\Omega\times J$-effect or the $\delta$ dynamo
effect \citep{rad69}. Their importance is supported by the numerical
simulations \citep{2008A&A...491..353K,2011A&A...533A.108S}. We use
the equation for $\eta_{ijk}^{(\delta)}$ which was suggested in P08
(also, see, \citealp{2004astro.ph..7375R}): 
\begin{equation}
\eta_{ijk}^{(\delta)}=3\eta_{T}C_{\delta}f_{4}^{(d)}\frac{\Omega_{j}}{\Omega}\left\{ \tilde{\varphi}_{7}^{(w)}\delta_{ik}+\tilde{\varphi}_{2}^{(w)}\frac{\overline{B}_{i}\overline{B}_{k}}{\overline{B}^{2}}\right\} ,\label{eq:delta}
\end{equation}
where, $C_{\delta}$ measures the strength of the $\Omega\times J$
effect, $\tilde{\varphi}_{2,7}^{(w)}\left(\beta\right)$ are normalized
versions of the magnetic quenching functions $\varphi_{2,7}^{(w)}$
given in P08. They are defined as follows, $\tilde{\varphi}_{2,7}^{(w)}\left(\beta\right)=\frac{5}{3}\varphi_{2,7}^{(w)}\left(\beta\right)$.
The functions $f_{\{1-11\}}^{(a,d)}$in Eqs (\ref{alp2d},\ref{eq:pump},\ref{eq:diff},
\ref{eq:delta}) depend on the Coriolis number. They can be found
in P08, as well.


\end{document}